\documentclass{ws-ijmpd}
\usepackage{epsfig,amssymb,sublabel,comment}

\renewcommand{\vec}[1]{{\bf{#1}}}

\begin{document}

\title{Cosmological observations in a modified theory of gravity (MOG)}

\author{J. W. Moffat$^{1,2}$ and V. T. Toth$^1$\\
$^1$Perimeter Institute for Theoretical Physics, Waterloo, Ontario N2L 2Y5, Canada\\
$^2$Department of Physics, University of Waterloo, Waterloo, Ontario N2L 3G1, Canada}

\maketitle

\begin{abstract}
{Our modified gravity theory (MOG) is a gravitational theory without exotic dark matter, based on an action principle. MOG has been used successfully to model astrophysical phenomena such as galaxy rotation curves, galaxy cluster masses, and lensing. MOG may also be able to account for cosmological observations. We assume that the MOG point source solution can be used to describe extended distributions of matter via an appropriately modified Poisson equation. We use this result to model perturbation growth in MOG and find that it agrees well with the observed matter power spectrum at present. As the resolution of the power spectrum improves with increasing survey size, however, significant differences emerge between the predictions of MOG and the standard $\Lambda$CDM model, as in the absence of exotic dark matter, oscillations of the power spectrum in MOG are not suppressed. We can also use MOG to model the acoustic power spectrum of the cosmic microwave background. A suitably adapted semi-analytical model offers a first indication that MOG may pass this test, and correctly model the peak of the acoustic spectrum.}
\end{abstract}


\section{Introduction}

The preferred model of cosmology today, the $\Lambda$CDM model, provides an excellent fit to cosmological observations, but at a substantial cost: according to this model, {\em about 95\% of the universe is either invisible or undetectable, or possibly both}\cite{Komatsu2008}. This fact provides a strong incentive to seek alternative explanations that can account for cosmological observations without resorting to dark matter or Einstein's cosmological constant.

For gravitational theories designed to challenge the $\Lambda$CDM model, the bar is set increasingly higher by recent discoveries. Not only do such theories have to explain successfully the velocity dispersions, rotational curves, and gravitational lensing of galaxies and galaxy clusters, the theories must also be in accord with cosmological observations, notably the acoustic power spectrum of the cosmic microwave background (CMB), the matter power spectrum of galaxies, and the recent observation of the luminosity-distance relationship of high-$z$ supernovae, which is seen as evidence for ``dark energy''.

Modified Gravity (MOG\cite{Moffat2006a}) has been used successfully to account for galaxy cluster masses\cite{Brownstein2006b}, the rotation curves of galaxies\cite{Brownstein2006a,Brownstein:2009gz}, velocity dispersions of satellite galaxies\cite{Moffat2007}, and globular clusters\cite{Moffat2007a}. It was also used to offer an explanation for the Bullet Cluster\cite{Brownstein2007} without resorting to nonbaryonic dark matter.

MOG may also be able to meet the challenge posed by cosmological observations. We investigate two sets of observations in particular: the matter power spectrum that describes the spatial distribution of galaxies in the Universe, and the acoustic spectrum of the cosmic microwave background (CMB) radiation.

In the next section, we review the key features of MOG. This is followed by sections presenting detailed calculations for the galaxy power spectrum and the acoustic power spectrum of the CMB. A concluding section summarizes our results and maps out future steps.

\section{Modified Gravity Theory}

Modified Gravity (MOG) is a fully relativistic theory of gravitation that is derived from a relativistic action principle\cite{Moffat2006a} involving scalar, tensor, and vector fields. MOG has evolved as a result of investigations of Nonsymmetric Gravity Theory (NGT\cite{Moffat1995}), and most recently, it has taken the form of Scalar-Tensor-Vector Gravity (STVG\cite{Moffat2006a}). In the weak field approximation, STVG, NGT, and Metric-Skew-Tensor Gravity (MSTG\cite{Moffat2005}) produce similar results.

\subsection{Scalar-Tensor-Vector Gravity}
\label{sec:STVG}

Our modified gravity theory is based on postulating the existence of a massive vector field, $\phi_\mu$. The choice of a massive vector field is motivated by our desire to introduce a {\em repulsive} modification of the law of gravitation at short range. The vector field is coupled universally to matter. The theory, therefore, has three constants: in addition to the gravitational constant $G$, we must also consider the coupling constant $\omega$ that determines the coupling strength between the $\phi_\mu$ field and matter, and a further constant $\mu$ that arises as a result of considering a vector field of non-zero mass, and controls the coupling range. The theory promotes $G$, $\mu$, and $\omega$ to scalar fields, hence they are allowed to run, resulting in the following action\cite{Moffat2006a,Moffat2007e}.
\begin{equation}
S=S_G+S_\phi+S_S+S_M,
\end{equation}
where
\begin{eqnarray}
S_G&=&-\frac{1}{16\pi}\int\frac{1}{G}\left(R+2\Lambda\right)\sqrt{-g}~d^4x,\\
S_\phi&=&-\int\omega\left[\frac{1}{4}B^{\mu\nu}B_{\mu\nu}-\frac{1}{2}\mu^2\phi_\mu\phi^\mu+V_\phi(\phi)\right]\sqrt{-g}~d^4x,\\
S_S&=&\int\frac{1}{G}\left[\frac{1}{2}g^{\mu\nu}\left(\frac{\nabla_\mu G\nabla_\nu G}{G^2}+\frac{\nabla_\mu\mu\nabla_\nu\mu}{\mu^2}-\nabla_\mu\omega\nabla_\nu\omega\right)\right.\nonumber\\
&&-\left.\frac{V_G(G)}{G^2}-\frac{V_\mu(\mu)}{\mu^2}-V_\omega(\omega)\right]\sqrt{-g}~d^4x.
\end{eqnarray}
Here, $S_M$ is the ``matter'' action, while $B_{\mu\nu}=\partial_\mu\phi_\nu-\partial_\nu\phi_\mu$, and $V_\phi(\phi)$, $V_G(G)$, $V_\omega(\omega)$, and $V_\mu(\mu)$ denote the self-interaction potentials associated with the vector field and the three scalar fields. The symbol $\nabla_\mu$ is used to denote covariant differentiation with respect to the metric $g^{\mu\nu}$, while the symbols $R$, $\Lambda$, and $g$ represent the Ricci-scalar, the cosmological constant, and the determinant of the metric tensor, respectively. We define the Ricci tensor as
\begin{equation} R_{\mu\nu}=\partial_\alpha\Gamma^\alpha_{\mu\nu}-\partial_\nu\Gamma^\alpha_{\mu\alpha}+\Gamma^\alpha_{\mu\nu}\Gamma^\beta_{\alpha\beta}-\Gamma^\alpha_{\mu\beta}\Gamma^\beta_{\alpha\nu}.
\end{equation}
Our units are such that the speed of light, $c=1$; we use the metric signature $(+,-,-,-)$.

The apparent ``wrong'' sign of the $\nabla_\mu\omega\nabla^\mu\omega$ term in the Lagrangian is of potential concern; however, we found that in all the solutions (including numerical solutions) considered to date, $\omega$ remains constant. Keeping $\omega$ as a dynamical scalar field (with the ``wrong'' sign in the Lagrangian) allowed us to develop a parameter-free solution\cite{Moffat2007e}, but we anticipate that the $\omega$ field may disappear from the theory as it is being further developed.

A direct numerical solution of the theory's field equations in the spatially homogeneous, isotropic case (FLRW cosmology) yields an expanding universe. Choosing a constant $V_G$ as one of the initial parameters of the solution, the age of the universe can be adjusted to fit observation. As an alternative, we also considered changing the overall sign of the kinetic terms in $S_S$; this solution, which violates several energy conditions, but keeps the energy density $\rho$ positive, is a ``bouncing'' cosmology (indeed, a classical bouncing cosmology requires that some or all of the energy conditions be violated). In this cosmology, the age of the universe since the bounce and the density of the universe at the time of the bounce can be tuned by choosing an appropriate constant $V_G$. Either way, a solution in which the age of the universe is in agreement with observation can be obtained. These solutions are a subject of further study, which will be reported elsewhere.

\subsection{Point particles in a spherically symmetric field}

For a point particle moving in the spherically symmetric field of a gravitating source, a particularly simple solution for the acceleration is obtained\cite{Moffat2008a}:
\begin{equation}
\ddot{r}=-\frac{G_NM}{r^2}\left[1+\alpha-\alpha(1+\mu r)e^{-\mu r}\right],
\label{eq:MOGaccel}
\end{equation}
where $M$ is the source mass, while $\alpha$ determines the strength of the ``fifth force'' interaction, and $\mu$ controls its range. In prior work, $\alpha$ and $\mu$ were considered free parameters that were fitted to data. Our recent work\cite{Moffat2007e} allows us to determine $\alpha$ and $\mu$ as functions of the source mass $M$:
\begin{equation}
\alpha=\frac{M}{(\sqrt{M}+E)^2}\left(\frac{G_\infty}{G_N}-1\right),
\label{eq:alpha2}
\end{equation}
and
\begin{equation}
\mu=\frac{D}{\sqrt{M}}.
\end{equation}

This solution can be seen to satisfy the field equations in the spherically symmetric case either numerically or by deriving an approximate solution analytically\cite{Moffat2007e}. The numerical values for $D$ and $E$ are determined by matching the result against galaxy rotation curves\cite{Moffat2007e}:
\begin{eqnarray}
D&\simeq&6250~M_\odot^{1/2}\mathrm{kpc}^{-1},\label{eq:C2}\\
E&\simeq&25000~M_\odot^{1/2}.\label{eq:C1}
\end{eqnarray}
The value of $G_\infty\simeq 20G_N$ is set to ensure that at the horizon distance, the effective strength of gravity is about 6 times $G_N$, eliminating the need for cold dark matter in cosmological calculations, as described in the previous section.

\subsection{The MOG Poisson Equation}
\label{sec:Poisson}

The acceleration law (\ref{eq:MOGaccel}) is associated with the potential,
\begin{equation}
\Phi=-\frac{G_\infty M}{r}\left[1-\frac{\alpha}{1+\alpha}e^{-\mu r}\right]=\Phi_N+\Phi_Y,
\label{eq:pot}
\end{equation}
where
\begin{equation}
\Phi_N=-\frac{G_\infty M}{r}
\label{eq:PhiN}
\end{equation}
is the Newtonian gravitational potential with $G_\infty=(1+\alpha)G_N$ as the gravitational constant, and
\begin{equation}
\Phi_Y=\frac{\alpha}{1+\alpha}G_\infty M\frac{e^{-\mu r}}{r}
\label{eq:PhiM}
\end{equation}
is the Yukawa-potential. These potentials are associated with the corresponding Poisson and inhomogeneous Helmholtz equations, which are given by Ref.~\refcite{Brownstein2007}:
\begin{align}
\nabla^2\Phi_N(\vec{r})&=4\pi G_\infty\rho(\vec{r}),\label{eq:PoissonN}\\
(\nabla^2-\mu^2)\Phi_Y(\vec{r})&=-4\pi\frac{\alpha}{1+\alpha}G_\infty\rho(\vec{r}).\label{eq:PoissonM}
\end{align}
Full solutions to these potentials are given by
\begin{align}
\Phi_N(\vec{r})&=-G_\infty\int{\frac{\rho(\vec{\tilde{r}})}{|\vec{r}-\vec{\tilde{r}}|}}~d^3\vec{\tilde{r}},\\
\Phi_Y(\vec{r})&=\frac{\alpha}{1+\alpha}G_\infty\int{\frac{e^{-\mu|\vec{r}-\vec{\tilde{r}}|}\rho(\vec{\tilde{r}})}{|\vec{r}-\vec{\tilde{r}}|}}~d^3\vec{\tilde{r}}.\label{eq:PsiY}
\end{align}
These solutions can be verified against Eqs.~(\ref{eq:PhiN}) and (\ref{eq:PhiM}) by applying the delta function point source density $\rho(\vec{r})=M\delta^3(\vec{r})$.

Strictly speaking, (\ref{eq:PsiY}) is a valid solution only when $\alpha$ is approximately constant. For inhomogeneous matter distributions, $\alpha$ is expected to vary as a function of matter density, and as such, this naive application of the spherically symmetric, static vacuum solution to model extended distributions of matter breaks down. However, for small perturbations of a homogeneous background, we expect (\ref{eq:PsiY}) to remain valid; this expectation can be verified once generalized (approximate) solutions of the theory in the presence of matter become available.

Combining Eq.~(\ref{eq:pot}) with Eqs.~(\ref{eq:PoissonN}) and (\ref{eq:PoissonM}) yields
\begin{eqnarray}
\label{eq:Poisson}
\nabla^2\Phi&=&4\pi G_N\rho(\vec{r})+\mu^2\Phi_Y(\vec{r})\\
&=&4\pi G_N\rho(\vec{r})+
\alpha\mu^2G_N\int{\frac{e^{-\mu|\vec{r}-\vec{\tilde{r}}|}\rho(\vec{\tilde{r}})}{|\vec{r}-\vec{\tilde{r}}|}}~d^3\vec{\tilde{r}},\nonumber
\end{eqnarray}
containing, in addition to the usual Newtonian term, a nonlocal source term on the right-hand side.

\section{MOG and the matter power spectrum}

The distribution of mass in the universe is not uniform. Due to gravitational self-attraction, matter tends to ``clump'' into ever denser concentrations, leaving large voids in between. In the early universe, this process is counteracted by pressure. The process is further complicated by the fact that in the early universe, the energy density of radiation was comparable to that of matter.

\subsection{Density fluctuations in Newtonian gravity}

To first order, this process can be investigated using perturbation theory. Taking an arbitrary initial distribution, one can proceed to introduce small perturbations in the density, velocity, and acceleration fields. These lead to a second-order differential equation for the density perturbation that can be solved analytically or numerically. This yields the {\em transfer function}, which determines how an initial density distribution evolves as a function of time in the presence of small perturbations.

\subsubsection{Newtonian theory of small fluctuations}

In order to see how this theory can be developed for MOG, we must first review how the density perturbation equation is derived in the Newtonian case. Our treatment follows closely the approach presented by Ref.~\refcite{Weinberg1972}. We begin with three equations: the continuity equation, the Euler equation, and the Poisson equation.
\begin{align}\sublabon{equation}
\frac{\partial\rho}{\partial t}+\nabla\cdot(\rho\vec{v})&=0,\\
\frac{\partial\vec{v}}{\partial t}+(\vec{v}\cdot\nabla)\vec{v}&=-\frac{1}{\rho}\nabla p+\vec{g},\\
\nabla\cdot\vec{g}&=-4\pi G\rho.
\end{align}\sublaboff{equation}

First, we perturb $\rho$, $p$, $\vec{v}$ and $\vec{g}$. Spelled out in full, we get:
\begin{align}\sublabon{equation}
\frac{\partial(\rho+\delta\rho)}{\partial t}+\nabla\cdot[(\rho+\delta\rho)(\vec{v}+\vec{\delta v})]&=0,\\
\frac{\partial(\vec{v}+\vec{\delta v})}{\partial t}+[(\vec{v}+\vec{\delta v})\cdot\nabla](\vec{v}+\vec{\delta v})
&=-\frac{1}{\rho+\delta\rho}\nabla(p+\delta p)+\vec{g}+\vec{\delta g},\\
\nabla\cdot(\vec{g}+\vec{\delta g})&=-4\pi G(\rho+\delta\rho).
\end{align}\sublaboff{equation}

Subtracting the original set of equations from the new set, using $1/(\rho+\delta\rho)=(\rho-\delta\rho)/[\rho^2-(\delta\rho)^2]=1/\rho-\delta\rho/\rho^2$, and eliminating second-order terms, we obtain
\begin{align}\sublabon{equation}
\frac{\partial\delta\rho}{\partial t}+\nabla\cdot(\delta\rho\vec{v}+\rho\vec{\delta v})&=0,\\
\frac{\partial\vec{\delta v}}{\partial t}+(\vec{v}\cdot\nabla)\vec{\delta v}+(\vec{\delta v}\cdot\nabla)\vec{v}&=\frac{\delta\rho}{\rho^2}\nabla p-\frac{1}{\rho}\nabla\delta p+\vec{\delta g},\\
\nabla\cdot\vec{\delta g}&=-4\pi G\delta\rho.
\end{align}\sublaboff{equation}

A further substitution can be made by observing that $\delta p=(\delta p/\delta\rho)\delta\rho=c_s^2\delta\rho$ where $c_s^2=(\partial p/\partial\rho)_\mathrm{adiabatic}$ is the speed of sound. We can also eliminate terms by observing that the original (unperturbed) state is spatially homogeneous, hence $\nabla\rho=\nabla p=0$:
\begin{align}\sublabon{equation}
\frac{\partial\delta\rho}{\partial t}+\vec{v}\cdot\nabla\delta\rho+\delta\rho\nabla\cdot\vec{v}+\rho\nabla\cdot\vec{\delta v}&=0,\\
\frac{\partial\vec{\delta v}}{\partial t}+(\vec{v}\cdot\nabla)\vec{\delta v}+(\vec{\delta v}\cdot\nabla)\vec{v}&=-\frac{c_s^2}{\rho}\nabla\delta\rho+\vec{\delta g},\\
\nabla\cdot\vec{\delta g}&=-4\pi G\delta\rho.
\end{align}\sublaboff{equation}
Now we note that $\vec{v}=H\vec{x}$, hence
\begin{align*}\nabla\cdot\vec{v}&=H\nabla\cdot\vec{x}=3H,\\
(\delta\vec{v}\cdot\nabla)\vec{v}&=(\delta\vec{v}\cdot\nabla)(H\vec{x})=H(\delta\vec{v}\cdot\nabla)\vec{x}=H\delta\vec{v}.
\end{align*}
Therefore,
\begin{align}\sublabon{equation}
\frac{\partial\delta\rho}{\partial t}+\vec{v}\cdot\nabla\delta\rho+3H\delta\rho+\rho\nabla\cdot\vec{\delta v}&=0,\\
\frac{\partial\vec{\delta v}}{\partial t}+(\vec{v}\cdot\nabla)\vec{\delta v}+H\vec{\delta v}&=-\frac{c_s^2}{\rho}\nabla\delta\rho+\vec{\delta g},\\
\nabla\cdot\vec{\delta g}&=-4\pi G\delta\rho.
\end{align}\sublaboff{equation}
The next step is a change of spatial coordinates to coordinates comoving with the Hubble flow:
\[\vec{x}=a(t)\vec{q}.\]
This means
\[\left(\frac{\partial}{\partial t}\right)_\vec{q}=\left(\frac{\partial}{\partial t}\right)_\vec{x}+\vec{v}\nabla_\vec{x},\]
and
\[\nabla_\vec{q}=a\nabla_\vec{x}.\]
After this change of coordinates, our system of equations becomes
\begin{align}\sublabon{equation}
\frac{\partial\delta\rho}{\partial t}+3H\delta\rho+\frac{1}{a}\rho\nabla\cdot\vec{\delta v}&=0\label{eq:7a},\\
\frac{\partial\vec{\delta v}}{\partial t}+H\vec{\delta v}&=-\frac{c_s^2}{a\rho}\nabla\delta\rho+\vec{\delta g}\label{eq:7b},\\
\nabla\cdot\vec{\delta g}=-4\pi aG\delta\rho.\label{eq:7c}
\end{align}\sublaboff{equation}
Now is the time to introduce the fractional amplitude $\delta=\delta\rho/\rho$. Dividing (\ref{eq:7a}) with $\rho$, we get
\begin{equation}
\dot\delta+\frac{\dot\rho}{\rho}\delta+3H\delta+\frac{1}{a}\nabla\cdot\vec{\delta v}=0.
\end{equation}
However, since $\rho=\rho_0a_0^3/a^3$, and hence $\dot\rho/\rho=-3\dot{a}/a$, the second and third terms cancel, to give
\begin{equation}
-a\dot\delta=\nabla\vec{\delta v}.
\end{equation}
Taking the gradient of (\ref{eq:7b}) and using (\ref{eq:7c}) to express $\nabla\cdot\vec{\delta g}$, we get
\begin{equation}
\frac{\partial}{\partial t}(-a\dot\delta)+H(-a\dot\delta)=-\frac{c_s^2}{a}\nabla^2\delta-4\pi Ga\rho\delta.
\end{equation}
Spelling out the derivatives, and dividing both sides with $a$, we obtain
\begin{equation}
\ddot\delta+2H\dot\delta-\frac{c_s^2}{a^2}\nabla^2\delta-4\pi G\rho\delta=0.\label{eq:delta}
\end{equation}
For every Fourier mode $\delta=\delta_\vec{k}(t)e^{i\vec{k}\cdot\vec{q}}$ (such that $\nabla^2\delta=-k^2\delta$), this gives
\begin{equation}
\ddot\delta_\vec{k}+2H\dot\delta_\vec{k}+\left(\frac{c_s^2k^2}{a^2}-4\pi G\rho\right)\delta_\vec{k}=0.
\label{eq:ddotdelta}
\end{equation}
The quantity $k/a$ is called the co-moving wave number.

If $k$ is large, solutions to (\ref{eq:ddotdelta}) are dominated by an oscillatory term; for small $k$, a growth term predominates.

A solution to (\ref{eq:ddotdelta}) tells us how a power spectrum evolves over time, as a function of the wave number; it does not specify the initial power spectrum. For this reason, solutions to (\ref{eq:ddotdelta}) are typically written in the form of a transfer function
\begin{equation}
T(k)=\frac{\delta_k(z=0)\delta_0(z=\infty)}{\delta_k(z=\infty)\delta_0(z=0)}.
\end{equation}

If the initial power spectrum and the transfer function are known, the power spectrum at a later time can be calculated (without accounting for small effects) as
\begin{equation}
P(k)=T^2(k)P_0(k).
\end{equation}

$P(k)$ is a dimensioned quantity. It is possible to form the dimensionless power spectrum
\begin{equation}
\Delta^2(k)=Ak^3T^2(k)P_0(k),
\end{equation}
where $A$ is a normalization constant determined by observation. This form often appears in the literature. In the present work, however, we are using $P(k)$ instead of $\Delta(k)$.

The initial power spectrum is believed to be a scale invariant power spectrum:
\begin{equation}
P_0(k)\propto k^n,
\end{equation}
where $n\simeq 1.$ A recent estimate\cite{Komatsu2008} on $n$ is $n=0.963^{+0.014}_{-0.015}$.

\subsubsection{Analytical approximation}

Eq.~(\ref{eq:ddotdelta}) is not difficult to solve in principle. The solution can be written as the sum of oscillatory and growing terms. The usual physical interpretation is that when pressure is sufficient to counteract gravitational attraction, this mechanism prevents the growth of density fluctuations, and their energy is dissipated instead in the form of sound waves. When the pressure is low, however, the growth term dominates and fluctuations grow. Put into the context of an expanding universe, one can conclude that in the early stages, when the universe was hot and dense, the oscillatory term had to dominate. Later, the growth term took over, the perturbation spectrum ``froze'', affected only by uniform growth afterwards.

In practice, several issues complicate the problem. First, the early universe cannot be modeled by matter alone; it contained a mix of matter and radiation (and, possibly, neutrinos and cold dark matter.) To correctly describe this case even using the linear perturbation theory outlined in the previous sections, one needs to resort to a system of coupled differential equations describing the different mediums. Second, if the perturbations are sufficiently strong, linear theory may no longer be valid. Third, other nonlinear effects, including Silk-damping\cite{Padmanabhan1993}, cannot be excluded as their contribution is significant (indeed, Silk damping at higher wave numbers is one of the reasons why a baryon-only cosmological model based on Einstein's theory of gravity without dark matter fails to account for the matter power spectrum.)

The authors of Ref.~\refcite{EH1998} addressed all these issues when they developed a semi-analytical solution to the baryon transfer function. This solution reportedly yields good results in the full range of $0\leq\Omega_b\leq1$. Furthermore, unlike other approximations and numerical software codes, this approach keeps the essential physics transparent, allowing us to adapt the formulation to the MOG case.

In Ref.~\refcite{EH1998} the transfer function is written as the sum of a baryonic term $T_b$ and a cold dark matter term $T_c$:
\begin{equation}
T(k)=\frac{\Omega_b}{\Omega_m}T_b(k)+\frac{\Omega_c}{\Omega_m}T_c(k),
\end{equation}
where $\Omega_c$ represents the cold dark matter content of the universe relative to the critical density. As we are investigating a cosmology with no cold dark matter, we ignore $T_c$. The baryonic part of the transfer function departs from the cold dark matter case on scales comparable to, or smaller than, the sound horizon. Consequently, the baryonic transfer function is written as
\begin{equation}
T_b(k)=\left[\frac{\tilde{T}_0(k,1,1)}{1+(ks/5.2)^2}+\frac{\alpha_be^{-(k/k_\mathrm{Silk})^{1.4}}}{1+(\beta_b/ks)^3}\right]\frac{\sin{k\tilde{s}}}{k\tilde{s}},
\label{eq:Tb}
\end{equation}
with
\begin{equation}
\tilde{T}_0(k,\alpha_c,\beta_c)=\frac{\ln{(e+1.8\beta_c\bar{q})}}{\ln{(e+1.8\beta_c\bar{q})}+C\bar{q}^2},
\end{equation}
where
\begin{equation}
C=\frac{14.2}{\alpha_c}+\frac{386}{1+69.9\bar{q}^{1.08}},
\end{equation}
and
\begin{equation}
\bar{q}=k\Theta_{2.7}^2\left(\Omega_mh^2\right)^{-1}.
\end{equation}
The sound horizon is calculated as
\begin{equation}
s=\frac{2}{3k_\mathrm{eq}}\sqrt{\frac{G}{R_\mathrm{eq}}}\ln{\frac{\sqrt{1+R_d}+\sqrt{R_d+R_\mathrm{eq}}}{1+\sqrt{R_\mathrm{eq}}}}.
\end{equation}
The scale at the equalization epoch is calculated as
\begin{equation}
k_\mathrm{eq}=7.46\times 10^{-2}\Omega_mh^2\Theta_{2.7}^{-2}.
\end{equation}
The transition from a radiation-dominated to a matter-dominated era happens at the redshift
\begin{equation}
z_\mathrm{eq}=25000\Omega_mh^2\Theta_{2.7}^{-4},
\end{equation}
while the drag era is defined as
\begin{equation}
z_d=1291\frac{(\Omega_mh^2)^{0.251}}{1+0.659(\Omega_mh^2)^{0.828}}[1+b_1(\Omega_mh^2)^{b_2}],
\end{equation}
where
\begin{equation}
b_1=0.313(\Omega_mh^2)^{-0.419}[1+0.607(\Omega_mh^2)^{0.674}],
\end{equation}
and
\begin{equation}
b_2=0.238(\Omega_mh^2)^{0.223}.
\end{equation}
The baryon-to-photon density ratio at a given redshift is calculated as
\begin{equation}
R=31.5\Omega_mh^2\Theta_{2.7}^{-4}\frac{1000}{z}.
\end{equation}
The Silk damping scale is obtained using
\begin{equation}
k_\mathrm{Silk}=1.6(\Omega_bh^2)^{0.52}(\Omega_mh^2)^{0.73}[1+(10.4\Omega_mh^2)^{-0.95}].
\label{eq:kSilk}
\end{equation}
The coefficients in the second term of the baryonic transfer function are written as
\begin{equation}
\alpha_b=2.07k_\mathrm{eq}s(1+R_d)^{-3/4}F\left(\frac{1+z_\mathrm{eq}}{1+z_d}\right),
\end{equation}
\begin{equation}
\beta_b=0.5+\frac{\Omega_b}{\Omega_m}+\left(3-2\frac{\Omega_b}{\Omega_m}\right)\sqrt{(17.2\Omega_mh^2)^2+1},
\end{equation}
where we used the function
\begin{equation}
F(y)=y\left[-6\sqrt{1+y}+(2+3y)\ln{\frac{\sqrt{1+y}+1}{\sqrt{1+y}-1}}\right].
\end{equation}
A shifting of nodes in the baryonic transfer function is accounted for by the quantity
\begin{equation}
\tilde{s}(k)=\frac{s}{\left[1+\left(\beta_\mathrm{node}/ks\right)^3\right]^{1/3}},
\end{equation}
where
\begin{equation}
\beta_\mathrm{node}=8.41(\Omega_mh^2)^{0.435}.
\end{equation}

The symbol $\Theta_{2.7}=T/2.7$ is the temperature of the CMB relative to 2.7~K, while $h=H/(100~\mathrm{km/s/Mpc})$. The wave number $k$ is in units of Mpc$^{-1}$.

\subsection{Density fluctuations in Modified Gravity}

\label{sec:MOGdensity}

\begin{figure}[t]
\begin{center}
\includegraphics[width=0.75\linewidth]{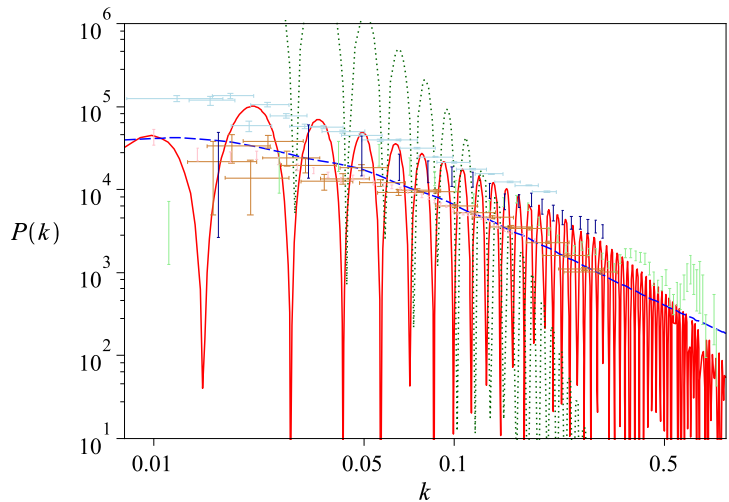}
\end{center}
\caption{The matter power spectrum. Three models are compared against five data sets (see text): $\Lambda$CDM (dashed blue line, $\Omega_b=0.035$, $\Omega_c=0.245$, $\Omega_\Lambda=0.72$, $H=71$~km/s/Mpc), a baryon-only model (dotted green line, $\Omega_b=0.035$, $H=71$~km/s/Mpc), and MOG (solid red line, $\alpha=19$, $\mu=5h$~Mpc$^{-1}$, $\Omega_b=0.035$, $H=71$~km/s/Mpc.) Data points are colored light blue (SDSS 2006), gold (SDSS 2004), pink (2dF), light green (UKST), and dark blue (CfA).}
\label{fig:Pk}
\end{figure}

We consider the MOG Poisson equation (\ref{eq:Poisson}), established in section~\ref{sec:Poisson}. As the initial unperturbed distribution is assumed to be homogeneous, $\rho$ is not a function of $\vec{r}$ and can be taken outside the integral sign:
\begin{equation}
\Phi_Y(\vec{r})=G_N\alpha\rho\int\frac{1}{|\vec{r}-\vec{r}'|}e^{-\mu|\vec{r}-\vec{r}'|}d^3\vec{r}'.
\end{equation}
Varying $\rho$, we get
\begin{eqnarray}
\nabla\cdot\delta\vec{g}(\vec{r})&=&-4\pi G_N\delta\rho(\vec{r})\nonumber\\
&&-\mu^2G_N\alpha\delta\rho\int\frac{1}{|\vec{r}-\vec{r}'|}e^{-\mu|\vec{r}-\vec{r}'|}d^3\vec{r}'.
\end{eqnarray}
Accordingly, (\ref{eq:delta}) now reads
\begin{equation}
\ddot\delta+2H\dot\delta-\frac{c_s^2}{a^2}\nabla^2\delta-4\pi G_N\rho\delta-\mu^2G_N\alpha\rho\delta\int\frac{e^{-\mu|\vec{r}-\vec{r}'|}}{|\vec{r}-\vec{r}'|}d^3\vec{r}'=0.
\label{eq:19}
\end{equation}
The integral can be readily calculated. Assuming that $|\vec{r}-\vec{r}'|$ runs from 0 to the comoving wavelength $a/k$, we get
\begin{eqnarray}
\int\frac{e^{-\mu|\vec{r}-\vec{r}'|}}{|\vec{r}-\vec{r}'|}d^3\vec{r}'&=&2\int\limits_0^{\pi/2}\int\limits_0^{2\pi}\int\limits_0^{a/k}\frac{e^{-\mu r}}{r}~r^2\sin{\theta}~dr~d\phi~d\theta\nonumber\\
&=&\frac{4\pi\left[1-(1+\mu a/k)e^{-\mu a/k}\right]}{\mu^2}.
\end{eqnarray}
Substituting into (\ref{eq:19}), we get
\begin{eqnarray}
\ddot\delta+2H\dot\delta-\frac{c_s^2}{a^2}\nabla^2\delta-4\pi G_N\rho\delta&&\nonumber\\
{}-4\pi G_N\alpha\left[1-\left(1+\frac{\mu a}{k}\right)e^{-\mu a/k}\right]\rho\delta&=&0,
\end{eqnarray}
or
\begin{eqnarray}
\ddot\delta+2H\dot\delta-\frac{c_s^2}{a^2}\nabla^2\delta&&\\
-4\pi G_N\left\{1+\alpha\left[1-\left(1+\frac{\mu a}{k}\right)e^{-\mu a/k}\right]\right\}\rho\delta&=&0.\nonumber
\end{eqnarray}

This demonstrates how the effective gravitational constant
\begin{equation}
G_\mathrm{eff}=G_N\left\{1+\alpha\left[1-\left(1+\frac{\mu a}{k}\right)e^{-\mu a/k}\right]\right\}
\end{equation}
depends on the wave number.

Using $G_\mathrm{eff}$, we can express the perturbation equation as
\begin{equation}
\ddot\delta_\vec{k}+2H\dot\delta_\vec{k}+\left(\frac{c_s^2k^2}{a^2}-4\pi G_\mathrm{eff}\rho\right)\delta_\vec{k}=0.
\label{eq:MOGdelta}
\end{equation}

As the wave number $k$ appears only in the source term
\[\left(\frac{c_s^2k^2}{a^2}-4\pi G_\mathrm{eff}\rho\right),\]
it is easy to see that any solution of (\ref{eq:ddotdelta}) is also a solution of (\ref{eq:MOGdelta}), provided that $k$ is replaced by $k'$ in accordance with the following prescription:
\begin{equation}
k'^2=k^2+4\pi a^2\left(\frac{G_\mathrm{eff}-G_N}{G_N}\right)\lambda_J^{-2},
\end{equation}
where $\lambda_J=\sqrt{c_s^2/G_N\rho}$ is the Jeans wavelength.

This shifting of the wave number applies to the growth term of the baryonic transfer function (\ref{eq:Tb}). However, as the sound horizon scale is not affected by changes in the effective gravitational constant, terms containing the product $ks$ must remain unchanged. Furthermore, the Silk damping scale must also change as a result of changing gravity; this change is proportional to the $3/4$th power of $G$, as demonstrated by\footnote{cf. Eq. (4.210) in Ref.~\refcite{Padmanabhan1993}; note that $\Omega h^2\propto G$.} Ref.~\refcite{Padmanabhan1993}, thus
\begin{equation}
k'_\mathrm{Silk}=k_\mathrm{Silk}\left(\frac{G_\mathrm{eff}}{G_N}\right)^{3/4},
\end{equation}
(note also Eq.~\ref{eq:kSilk}). Using these considerations, we obtain the modified baryonic transfer function
\begin{equation}
T'_b(k)=\frac{\sin{k\tilde{s}}}{k\tilde{s}}\left\{\frac{\tilde{T}_0(k',1,1)}{1+(ks/5.2)^2}+\frac{\alpha_b\exp\left({-[k/k'_\mathrm{Silk}]^{1.4}}\right)}{1+(\beta_b/ks)^3}\right\}.
\end{equation}

The effects of these changes can be summed up as follows. At low values of $k$, the transfer function is suppressed. At high values of $k$, where the transfer function is usually suppressed by Silk damping, the effect of this suppression is reduced. The combined result is that the tilt of the transfer function changes, such that its peaks are now approximately in agreement with data points, as seen in Figure~\ref{fig:Pk}.

Data points shown in this figure come from several sources. First and foremost, the two data releases of the Sloan Digital Sky Survey (SDSS\cite{SDSS2004,SDSS2006}) are presented. Additionally, data from the 2dF Galaxy Redshift Survey\cite{2dF2006}, UKST\cite{UKST1999}, and CfA130\cite{CfA1994} surveys are shown. Apart from normalization issues, the data from these surveys are consistent in the range of $0.01~h~\mathrm{Mpc}^{-1}\leq k\leq 0.5~h~\mathrm{Mpc}^{-1}$. Some surveys provide data points outside this range, but they are not in agreement with each other.

\subsection{Discussion}

As a result of the combined effects of dampened structure growth at low values of $k$ and reduced Silk damping at high values of $k$, the slope of the MOG transfer function differs significantly from the slope of the baryonic transfer function, and matches closely the observed values of the matter power spectrum. On the other hand, the predictions of MOG and $\Lambda$CDM cosmology differ in fundamental ways.

\begin{figure}
\begin{center}
\includegraphics[width=0.75\linewidth]{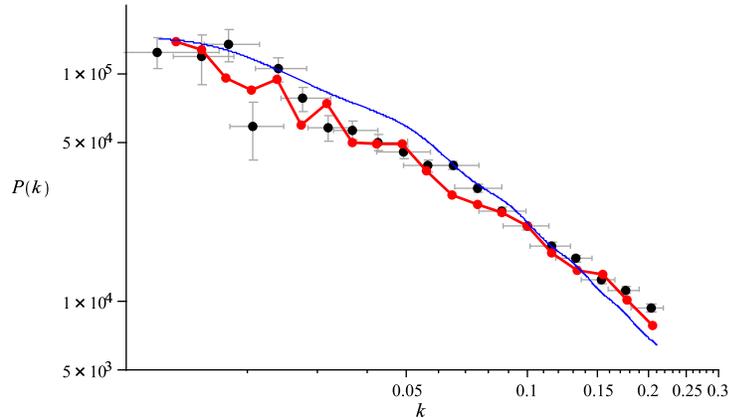}
\end{center}
\caption{The effect of window functions on the power spectrum is demonstrated by applying the SDSS luminous red galaxy survey window functions to the MOG prediction. Baryonic oscillations are greatly dampened in the resulting curve (solid red line). A normalized linear $\Lambda$CDM estimate is also shown (thin blue line) for comparison.}
\label{fig:baryon}
\end{figure}

First, MOG predicts oscillations in the power spectrum, which are not smoothed out by dark matter. These oscillations may be detectable in future galaxy surveys that utilize a large enough number of galaxies, and sufficiently narrow window functions in order to be sensitive to such fluctuations. However, the finite size of samples and the associated window functions used to produce presently available power spectra mask any such oscillations. To illustrate this, we applied the same window function to the MOG prediction, which resulted in a smoothed curve, seen in Figure~\ref{fig:baryon}. A $\chi^2$ comparison actually suggests that MOG offers a better fit ($\chi^2_\mathrm{MOG}=0.03$, $\chi^2_{\Lambda\mathrm{CDM}}=0.09$ per degrees of freedom), although we must be cautious: the $\Lambda$CDM approximation we used is not necessarily the best approximation available, and the MOG result is dependent on the validity of the analysis presented in this section, which was developed without the benefit of an interior solution.

Second, MOG predicts a dampened power spectrum at both high and low values of $k$ relative to $\Lambda$CDM. Observations at sufficiently high values of $k$ may not be practical, as we are entering sub-galactic length scales. Low values of $k$ are a different matter: as accurate three-dimensional information becomes available on ever more distant galaxies, power spectrum observations are likely to be extended in this direction.

In the present work, we made no attempt to account for the possibility of a non-zero neutrino mass, and its effects on the power spectrum. Given the uncertainties in the semi-analytical approximations that we utilized, such an attempt would not have been very fruitful. Future numerical work, however, must take into account the possibility of a non-negligible contribution of neutrinos to the matter density.

~\par

\section{MOG and the CMB}

The cosmic microwave background (CMB) is highly isotropic, showing only small temperature fluctuations as a function of sky direction. These fluctuations are not uniformly random; they show a distinct dependence on angular size, as has been demonstrated by the measurements of the Boomerang experiment\cite{Jones2006} and the Wilkinson Microwave Anisotropy Probe (WMAP\cite{Komatsu2008}).

The angular power spectrum of the CMB can be calculated in a variety of ways. The preferred method is to use numerical software, such as {\tt CMBFAST}\cite{Seljak1996}. Unfortunately, such software packages cannot easily be adapted for use with MOG. Instead, at the present time we opt to use the excellent semi-analytical approximation developed by Ref.~\refcite{Mukhanov2005}. While not as accurate as numerical software, it lends itself more easily to nontrivial modifications, as the physics remains evident in the equations.

What justifies the use of this semi-analytical approach is the fact that the phenomenology of MOG vs. dark matter can be understood easily. Collisionless cold dark matter interacts with normal matter only through gravity. In the late universe, the ratio of cold dark matter vs. baryonic matter varies significantly from region to region; this is why the results of the previous section are nontrivial and significant. However, in the early universe (recombination era), the universe was still largely homogeneous, and cold dark matter effectively acted as a ``gravity enhancer'': its effects can be mimicked by simply increasing the effective gravitational constant.

This may seem surprising in view of studies that have placed stringent constraints on the variability of $G$. For example, after substituting $G\rightarrow \lambda^2G$ in the Friedmann equation, the authors of Ref.~\refcite{Zahn2003} have shown that $\lambda$ is constrained to 10\% or better by WMAP data. At first sight, this seems inconsistent with our assertion that MOG, with $G_\mathrm{eff}>G_N$, can successfully mimic the effects of dark matter on the CMB acoustic spectrum. Yet that this is the case can be seen if one writes down the Friedmann equation after incorporating $\lambda$:
$$
H^2\sim\frac{8\pi}{3}\lambda^2G\rho.
$$
The full form of the substitution rule, therefore, is $G\rho\rightarrow\lambda^2G\rho$. In MOG, we substitute $G\rightarrow G_\mathrm{eff}$ and $\rho\rightarrow\rho_b$ (no CDM component), but $(G\rho)_{\Lambda\mathrm{CDM}}=(G_\mathrm{eff}\rho_b)_\mathrm{MOG}$, hence $\lambda\equiv 1$.

This discussion leads to a simple substitution rule that is applicable when the universe is approximately homogeneous. When a quantity containing $G$ appears in an equation describing a gravitational interaction, $G_\mathrm{eff}$ must be used. However, when a quantity like $\Omega_b$ is used to describe a nongravitational effect, the Newtonian value of $G_N$ must be retained.

Our choice to use Mukhanov's semianalytical approximation is motivated by the fact that these substitutions can be made in the formulae in a straightforward and unambiguous manner.

\subsection{Semi-analytical estimation of CMB anisotropies}

In Ref.~\refcite{Mukhanov2005} we find a calculation of the correlation function $C(l)$, where $l$ is the multipole number, of the acoustic power spectrum of the CMB using the solution
\begin{equation}
\frac{C(l)}{[C(l)]_{\mathrm{low}~l}}=\frac{100}{9}(O+N),
\end{equation}
where $l\gg 1$, $O$ denotes the oscillating part of the spectrum, while the non-oscillating part is written as the sum of three parts:
\begin{equation}
N=N_1+N_2+N_3.
\end{equation}
These, in turn, are expressed as
\begin{equation}
N_1=0.063\xi^2\frac{[P-0.22(l/l_f)^{0.3}-2.6]^2}{1+0.65(l/l_f)^{1.4}}e^{-(l/l_f)^2},
\end{equation}
\begin{equation}
N_2=\frac{0.037}{(1+\xi)^{1/2}}\frac{[P-0.22(l/l_s)^{0.3}+1.7]^2}{1+0.65(l/l_s)^{1.4}}e^{-(l/l_s)^2},
\end{equation}
\begin{equation}
N_3=\frac{0.033}{(1+\xi)^{3/2}}\frac{[P-0.5(l/l_s)^{0.55}+2.2]^2}{1+2(l/l_s)^2}e^{-(l/l_s)^2}.
\end{equation}
The oscillating part of the spectrum is written as
\begin{eqnarray}
O&=&e^{-(l/l_s)^2}\sqrt{\frac{\pi}{\bar\rho l}}\nonumber\\
&\times&\left[A_1\cos{\left(\bar\rho l+\frac{\pi}{4}\right)}+A_2\cos{\left(2\bar\rho l+\frac{\pi}{4}\right)}\right],
\end{eqnarray}
where
\begin{equation}
A_1=0.1\xi\frac{(P-0.78)^2-4.3}{(1+\xi)^{1/4}}e^{\frac{1}{2}(l_s^{-2}-l_f^{-2})l^2},
\end{equation}
and
\begin{equation}
A_2=0.14\frac{(0.5+0.36P)^2}{(1+\xi)^{1/2}}.
\end{equation}
The parameters that occur in these expressions are as follows. First, the baryon density parameter:
\begin{equation}
\xi=17\left(\Omega_bh_{75}^2\right),
\label{eq:xi}
\end{equation}
where $\Omega_b\simeq 0.035$ is the baryon content of the universe at present relative to the critical density, and $h_{75}=H/(75~\mathrm{km/s/Mpc})$. The growth term of the transfer function is represented by
\begin{equation}
P=\ln{\frac{\Omega_m^{-0.09}l}{200\sqrt{\Omega_mh_{75}^2}}},
\end{equation}
where $\Omega_m\simeq 0.3$ is the total matter content (baryonic matter, neutrinos, and cold dark matter). The free-streaming and Silk damping scales are determined, respectively, by
\begin{equation}
l_f=1300\left[1+7.8\times 10^{-2}\left(\Omega_mh_{75}^2\right)^{-1}\right]^{1/2}\Omega_m^{0.09},
\label{eq:lf}
\end{equation}
\begin{equation}
l_s=\frac{0.7l_f}{\sqrt{\frac{1+0.56\xi}{1+\xi}+\frac{0.8}{\xi(1+\xi)}\frac{\left(\Omega_mh_{75}^2\right)^{1/2}}{\left[1+\left(1+\frac{100}{7.8}\Omega_mh_{75}^2\right)^{-1/2}\right]^2}}}.
\end{equation}
Lastly, the location of the acoustic peaks is determined by the parameter\footnote{Note that we slightly adjusted the coefficients of (\ref{eq:lf}) and (\ref{eq:rho}), which improved the fit noticeably, while remaining fully consistent with Mukhanov's derivation.}
\begin{equation}
\bar\rho=0.015(1+0.13\xi)^{-1}(\Omega_mh_{75}^{3.1})^{0.16}.
\label{eq:rho}
\end{equation}

\subsection{The MOG CMB spectrum}

\begin{figure}[t]
\begin{center}
\includegraphics[width=0.75\linewidth]{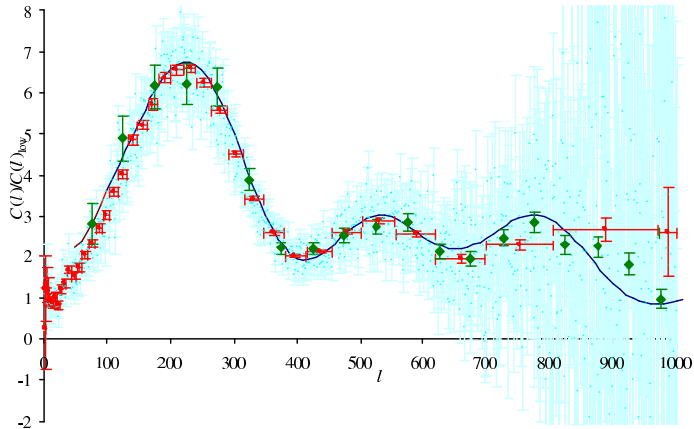}
\end{center}
\caption{MOG and the acoustic power spectrum. Calculated using $\Omega_M=0.3$, $\Omega_b=0.035$, $H_0=71$~km/s/Mpc. Also shown are the raw WMAP 3-year data set (light blue), binned averages with horizontal and vertical error bars provided by the WMAP project (red), and data from the Boomerang experiment (green).}
\label{fig:CMB}
\end{figure}

The semi-analytical approximation presented in the previous section can be adapted to the MOG case by making two important observations.

First, in all expressions involving the value of Mukhanov's $\Omega_m$ (which includes contributions from baryonic matter and cold dark matter using Newton's gravitational constant), we need to use $\Omega_M\simeq 0.3$ (which includes baryonic matter only, using the running value of the gravitational constant, $G_\mathrm{eff}\simeq 6G_N$). Second, we notice that the value of $\Omega_b$ in (\ref{eq:xi}) {\em does not depend on the effective value of the gravitational constant}, as this value is a function of the speed of sound, which depends on the (baryonic) matter density, regardless of gravitation. In other words, $\Omega_b\simeq 0.035$ is calculated using Newton's gravitational constant.

After we modify Mukhanov's semi-analytical formulation by taking these considerations into account, we obtain the fit to the acoustic power spectrum shown in Figure~\ref{fig:CMB}.

\subsection{Discussion}

As Figure~\ref{fig:CMB} demonstrates, to the extent that Mukhanov's formulation is applicable to MOG, the theory achieves agreement with the observed acoustic power spectrum. This result was obtained without fine-tuning or parameter fitting. The MOG constant $\mu$ was assumed to be equal to the inverse of the radius of the visible universe. Thereafter, the value of $\alpha$ is fixed if we wish to ensure $\Omega_M\simeq 0.3$. This was sufficient to achieve consistency with the data.

\section{Conclusions}

In this paper, we demonstrated how MOG can account for key cosmological observations using a minimum number of free parameters. We applied the MOG point source solution in a suitably modified form of the Poisson equation and re-derived the equations of structure growth. We found that the result is in agreement with presently available observational data.

Notably, we also found that as the available data sets grow in size, a significant, and likely irreconcilable disagreement emerges between the predictions of MOG and those of the $\Lambda$CDM concordance model. In $\Lambda$CDM, the presence of collisionless exotic dark matter leads to a significant dampening of the baryonic oscillations in the matter power spectrum: unit oscillations are suppressed, and appear only as a slight modulation of the power spectrum at shorter wavelengths. In contrast, unit oscillations are {\it not} suppressed in MOG. Presently, these oscillations are not seen only because the resolution of the data is not high enough: when we apply the appropriate bin sizes and window functions to a simulated data set, the resulting curve is nearly smooth. As galaxy surveys grow in size, however, bin sizes will get smaller and, if MOG is correct, the unit oscillations will emerge in the data.

We also investigated the acoustic power spectrum of the cosmic microwave background using MOG. Existing software codes, notably the program {\tt CMBFAST}\cite{Seljak1996} and its derivatives, are ill suited for this investigation as it is difficult to disentangle the use of quantities proportional to $G\rho$ in gravitational vs. nongravitational contexts. Before embarking on what seems to be a formidable task, we turned to a semi-analytical approximation\cite{Mukhanov2005}. While many of the approximations employed by Ref.~\refcite{Mukhanov2005} are not physically motivated, but numerical fitting formulae, nonetheless the role played by quantities proportional to $G\rho$ can be clearly discerned, and the formulae can be suitably adapted. While we recognize that this is not a conclusive result, we find it nonetheless encouraging that the CMB acoustic power spectrum was faithfully reproduced.

In conclusion, we have demonstrated that cosmological observations of the matter power spectrum and the CMB acoustic spectrum do not trivially rule out MOG as a possible alternative to the standard $\Lambda$CDM model of cosmology.

\section*{Acknowledgments}

The research was partially supported by National Research Council of Canada. Research at the Perimeter Institute for Theoretical Physics is supported by the Government of Canada through NSERC and by the Province of Ontario through the Ministry of Research and Innovation (MRI).

\bibliography{refs}
\bibliographystyle{unsrt}

\end{document}